\documentclass[twocolumn,prl,showpacs]{revtex4}

\usepackage{graphicx}
\usepackage{dcolumn}
\usepackage{float}
\usepackage{amsmath}

\newcommand{\comment}[1]{}

\begin{document}

\title{Optical spectroscopy of superconducting Ba$_{0.55}$K$_{0.45}$Fe$_2$As$_2$: evidence for strong coupling to low energy bosons.}

\author{J. Yang$^{1}$}
\author{D. H\"uvonen$^{2}$}
\author{U. Nagel$^{2}$}
\author{T. R$\tilde{o}\tilde{o}$m$^{2}$}
\author{N. Ni$^{3}$}
\author{P. C. Canfield$^{3}$}
\author{S. L. Bud'ko$^{3}$}
\author{J.P. Carbotte$^{1,4}$}
\author{T. Timusk,$^{1,4}$}

\email{timusk@mcmaster.ca.} 

\affiliation{$^{1}$Department of Physics and Astronomy, McMaster University, Hamilton, ON
L8S 4M1, Canada} 
\affiliation{$^{2}$National Institute of Chemical Physics and Biophysics, Akadeemia tee 23, 12618 Tallinn, Estonia}
\affiliation{$^{3}$Ames Laboratory and Department of Physics and Astronomy, Iowa State University, Ames Iowa 50011, USA.}
\affiliation{$^{4}$The Canadian Institute of Advanced Research, Toronto, Ontario M5G 1Z8, Canada.}

\date{\today}

\begin{abstract}
Optical spectroscopy on single crystals of the new iron arsenide superconductor Ba$_{0.55}$K$_{0.45}$Fe$_2$As$_2$ shows that the infrared spectrum consists of two major components: a strong metallic Drude band and a well separated mid infrared absorption centered at 0.7 eV.  It is difficult to separate the two components unambiguously but several fits of Lorentzian peaks suggest a model with a Drude peak having a plasma frequency of 1.8 to 2.1 eV and a midinfrared peak with a plasma frequency of 2.5 eV.  In contrast to the cuprate superconductors the scattering rate obtained from the extended Drude model saturates at 150  meV as compared to 500 meV for a typical cuprate.  Detailed analysis of the frequency dependent scattering rate shows that the charge carriers interact with broad bosonic spectrum with a peak at 25 meV and a coupling constant $\lambda \approx 2$ at low temperature. As the temperature increases this coupling weakens to $\lambda=0.6$ at ambient temperature.  This suggests a bosonic spectrum that is similar to what is seen in the lower $T_c$ cuprates.
\end{abstract}


\maketitle

With the discovery of superconductivity in the iron arsenic compounds
with $T_c$'s up to 54K\cite{kamihara08,takahashi08,chen08gf,ren08,chen08ty,wang08,rotter08,ni08}, an important question relates to the nature of the
driving mechanism in this class of materials.  For conventional
superconductors the pairing glue is provided by the electron-phonon
interaction and the energy gap has s-wave symmetry. Even MgB$_2$ with a $T_c=40$ K
falls in this same class but has the additional feature that two
bands with distinct gap values are involved\cite{tsuda03,kristoffel02,nicol05}. By contrast the gap in the
cuprates has d-wave symmetry and there is a rapidly expanding body of
evidence that the mechanism is the exchange of antiferromagnetic spin
fluctuations\cite{chubukov03,carbotte99,abanov99a,zasadzinski06,norman98,johnson01}.  Credible alternate explanations have also been put
forward.There exist other classes of superconductors, one example are the ruthenates which
are chiral with a p-wave gap\cite{ishida98,luke98,mackenzie03}.

Spectroscopic techniques have proved very useful in determining
the physical mechanism of superconductivity. In the classic superconductors the spectroscopy of choice
has been tunneling. Analysis of current-voltage data within an
Eliashberg formalism has yielded a detailed picture of the
electron-phonon spectral function $\alpha^2F(\Omega)$\cite{mcmillan66,carbotte90}, which agrees well with first principles calculations and also with similar data obtained from
optics\cite{farnworth76,bock07}. More recently similar evidence, mainly from infrared
spectroscopy, has been obtained for the cuprates\cite{hwang07b,hwang08}. A complication that
arises for this case, is that even in the simplest approaches, the
electron boson spectral density involved can be different in the d-
(pairing) and the s- (renormalization) channel which is a source of some
uncertainty. Nevertheless the preponerence of evidence is that this
function involves the local spin susceptibility and points to a spin
exchange mechanism. From this perspective the iron arsenic  materials are
particularly interesting. Recent density functional calculations have
revealed low electron densities, a high density of states at the Fermi surface  and an
attendent tendency towards itinerent magnetism\cite{singh08,delacruz08,mazin08}. At the same time
calculations of $\alpha^2F(\Omega)$ have yielded a small value of the electron-phonon
mass enhancement leading to doubt that these are electron-phonon
superconductors\cite{boeri08}. Yet experiments indicate the gap has s-wave symmetry
possibly involving several bands\cite{liu08,zhao08,ding08}. In view of these facts it is
particularly important to see what light optical data can shed on the
central question of pairing glue in these systems.

To extract the optical constants we used standard  optical reflectance techniques over a wide range of frequencies ranging in  photon energy from 5 meV to 5 eV at temperatures from 28 K to 295 K\cite{homes93a}. Additional experiments were done in an immersion dewar between 5 K and 75 K.  The Ba$_{0.55}$K$_{0.45}$Fe$_2$As$_2$ crystals were grown with a flux technique and  had a nominal $T_c$ of 28 K\cite{ni08}.  Figure 1 shows the measured reflectance as a function of photon energy in meV at nine temperatures.  At low energies the reflectance varies approximately linearly with frequency but there  are features at 20 and 40 meV that get weaker as the temperature is raised.  As a function of temperature there is a monotonic decrease of reflectance but there is a notable rapid change in the 80 K region as compared to the regions above and below this temperature. The inset shows the ambient temperature reflectance over a wider range of energies and a comparison with Bi$_2$Sr$_2$CaCu$_2$O$_{8+\delta}$ a familiar high temperature superconductor.

% Figure 1 Reflectance
\begin{figure}[t]
 \vspace*{-0.5 cm}
 \includegraphics[width=8cm]{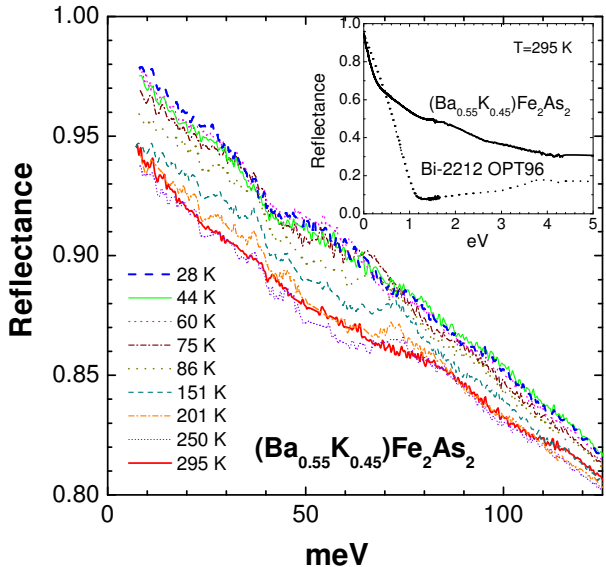}
%  \vspace*{-0.5 cm}%
\caption{(color online).The reflectance of Ba$_{1-x}$K$_x$Fe$_2$As$_2$  as a function of incident photon energy at nine temperatures.  The inset shows the reflectance at room temperature in a wider range of energies.  There is a sharp break in the monotonic linear decay of reflectance at 0.5 eV signaling the presence of a strong mid infrared band. Also shown is the reflectance of a typical high temperature superconductor.  The break at 0.5 eV  is absent in this material and the the strong linear decay of reflectance persists up to 1.2 eV. }
 \label{Fig1}
\end{figure}

The measured reflectance was converted to an optical conductivity by Kramers-Kronig analysis.  At low frequency Hagen-Rubens behavior was used and at high frequency, beyond 5 eV, free electron response was assumed.  Figure 2 shows the optical conductivity at ambient temperature and  at 28 K close to the superconducting transition temperature. We note a well-defined mid infrared peak at 0.7 eV and a separate Drude conductivity at low frequency, evidence of the metallic character of this material.  As the temperature is lowered the Drude band narrows and there appears to be a shift of spectral weight towards higher frequencies of the mid infrared band.  To be able to do further analysis of the optical spectra we need to separate the Drude band from the mid infrared band but it is clear from Figure 2 that this process will be difficult.  To illustrate the problem we have chosen two models for the optical conductivity.  Both split the absorption into three components,  the Drude conductivity, and two finite frequency Lorentz oscillators.   To estimate the influence of the mid infrared band on the Drude conductivity we have used two models, called A and B with different widths for the mid infrared band as shown in Fig. 2.  The model with the broader mid infrared band, model A, gives a narrower Drude band and vice versa for model B. We calculate the plasma frequencies for the Drude band to be 1.8 eV and 2.1 eV for models A and B respectively.  

% Figure 2 The conductivity
\begin{figure}[t]
%  \vspace*{-0.5 cm}
 \includegraphics[width=8cm]{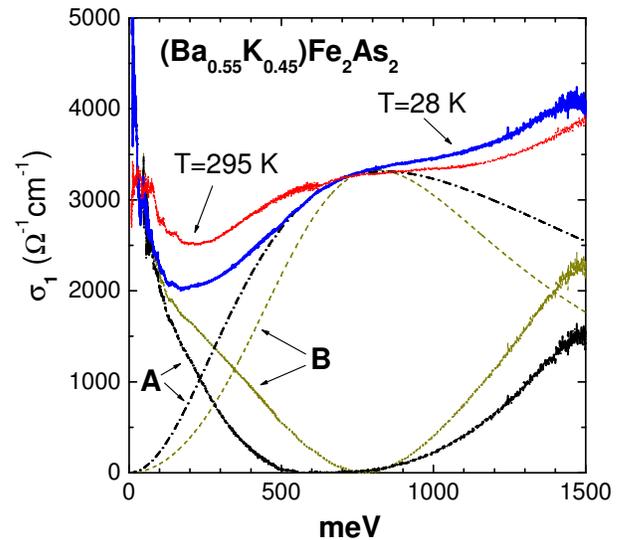}
%  \vspace*{-0.5 cm}%
\caption{(color online): The optical conductivity as a function of photon energy.  The conductivity is shown at two temperatures 28 K and 295 K.  Three components can be seen, a Drude band below 200 meV, a mid infrared band centered at  700 meV and a high frequency band above 1 eV.  The curves A and B represent two models for the separation of the Drude and the mid infrared bands.}
 \label{Fig2}
\end{figure}

Using the models A and B we can apply the extended Drude model to estimate the frequency dependent scattering rate for the charge carriers.  This scattering rate is given by a generalized Drude formula \begin{equation}
\sigma(T,\omega)={{\Omega_p^2} \over {4\pi}}{1 \over
{1/\tau(\omega)-i\omega(1+\lambda(\omega))}}
\end{equation}
where $\Omega_p^2=4\pi n e^2/m$ is the plasma frequency squared. The optical scattering
rate $1/\tau(\omega)={ne^2 \over m}{\cal R}{e}({1} / \sigma(\omega))$ and the optical mass enhancement factor $\lambda$ is given by $-\omega(1+\lambda(\omega))={ne^2 \over m} {\cal I}m(1 / \sigma(\omega))$.

% Fiure 3 The scattering rate and the Bosonic spectrum
\begin{figure}[t]
\vspace*{-0.5 cm}
\includegraphics[width=9cm]{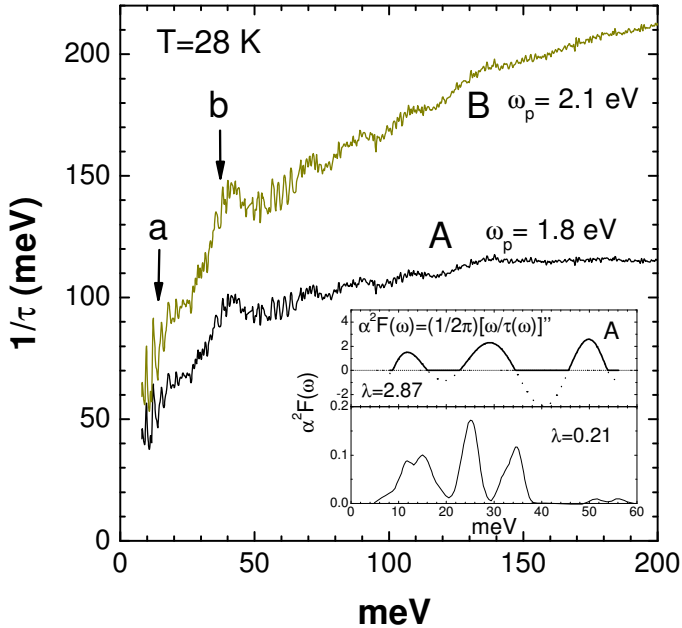}
\vspace*{-0.5 cm}%
\caption{(color online). Scattering rate calculated for two model A and B.  In both cases the scattering rate rises uniformly to saturate in the mid infrared region.  The fine structure marked with arrows at a and b may be caused by phonons.  An analysis within a phonon framework yields the bosonic spectrum shown in the inset and compared with a recent calculation of the same quantity by Boeri {\it et al.}}.
\label{Fig3}
\end{figure}

Figure 3 shows the scattering rate calculated for  models A and B for the mid infrared band.  Both curves rise smoothly to saturate in the mid infrared region at 200 meV for model A and 300 meV for model B.  There is some fine structure  shown with arrows a and b.  These peaks can be interpreted as onsets of scattering due to bosonic excitations at these frequencies.  An analysis in terms of the second derivative of the scattering rate\cite{marsiglio88,abanov01} allows one to calculate the bosonic spectral function from the scattering rate data. The result of such a calculation is shown in the inset to Figure 3 along with a phonon spectrum calculated by Boeri {\it et. al.}\cite{boeri08}.  However the very large value of the coupling constant $\lambda=2.87$ that this analysis yields is inconsistent with the temperature dependence of both the dc resistivity and the low frequency optical conductivity to be discussed below.  The high temperature behavior requires a reduction of $\lambda$ to $\lambda=0.5$ but such a temperature dependent coupling constant is incompatible with a  phonon mechanism.  We will therefore ignore the fine structure in the scattering rate and attribute it to experimental noise or a direct absorption by transverse optical phonons.  Instead, we will focus on the broader structure and attempt to fit the data with the MMP model of spin fluctuation scattering. 

% Figure 4 MMP fits to the spectrum 
\begin{figure}[t]
\vspace*{-0.5 cm}
\includegraphics[width=9cm]{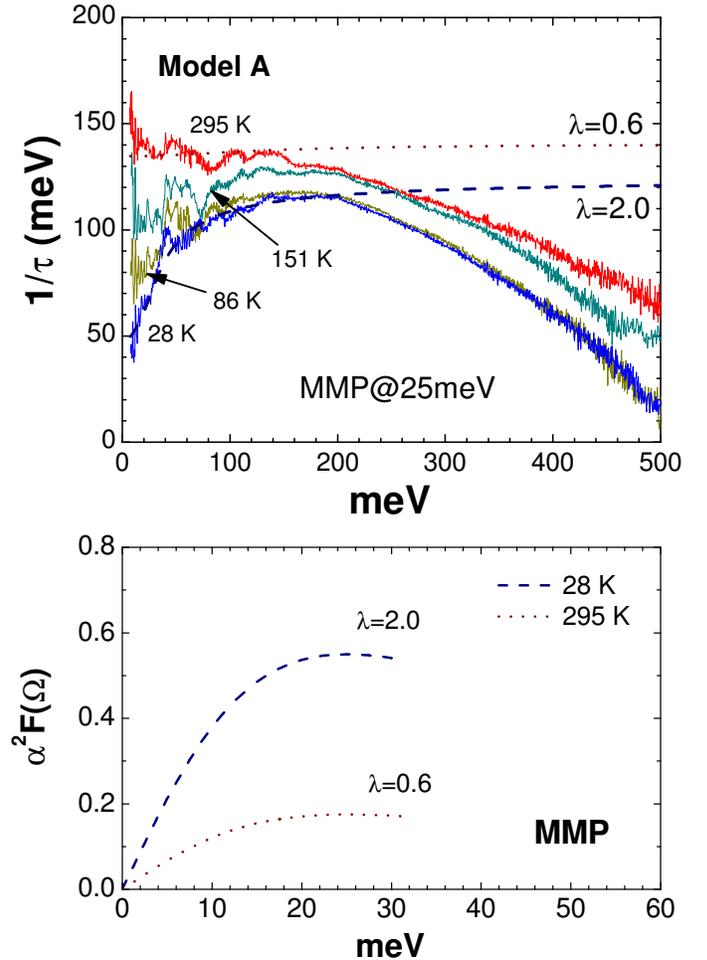}
\vspace*{-0.5 cm}%
\caption{(color online). Upper panel: scattering rate calculated for  model A  compared with fits to a magnetic fluctuation spectrum.  Lower panel bosonic spectral functions used to calculate the scattering rate at low and ambient temperatures.}
\label{Fig4}
\end{figure}

We choose a bosonic excitation spectrum based on the spin fluctuation model of Millis {\it et al.}\cite{millis90} the so called MMP spectrum used to model the magnetic excitations in the cuprates.  The lower panel shows the spectral function, a modified Lorentzian peak at 25 meV.  To fit the data we have to choose a dramatically different strength of the coupling constant $\lambda$ at different temperatures, $\lambda=0.6$ at high and  $\lambda=2.0$ low temperature.  While for a phonon spectrum this is not a valid choice, for magnetic fluctuations a temperature dependent spectrum is the norm.  This has been observed both with neutron scattering\cite{stock05,vignolle07} and optical spectroscopy\cite{hwang07b,hwang08} in the cuprates.  Several theoretical models of spin fluctuations predict the formation of magnetic excitons as the temperature is lowered and a peak develops in the local susceptibility\cite{abanov99,prelovsek06}.    The top panel in Figure 4 shows the resulting fits using an expression from Shulga\cite{shulga91} to calculate the temperature dependent scattering rate.  As the top panel shows we can fit the magnitude of the scattering and the low frequency dependence with such a model but at higher frequencies the experimental curves shift downward dramatically.  This may well be a finite band effect.

An independent test of our bosonic model is the temperature dependence of the dc resistivity.  Using our value of the plasma frequency we can calculate the scattering rate as function of temperature and estimate $\lambda$ from the formula: $1/\tau=2\pi\lambda T$.  There are several published dc resistivity curves\cite{rotter08,ni08,chen08gf2} and they all show a curve that flattens substantially between low temperature and ambient temperature.  Using a plasma frequency of 1.8 eV we find that $\lambda$ decreases from 1.8 to 0.5 on going from the 50 to 100 K range to the 250 to 300 K range for all dc measurements. Our own optical data can be analyzed in the low frequency limit to yield an optical resistivity.  This is in good agreement with the data of Chen {\it et al.}\cite{chen08gf2}.

In summary we find that optical spectroscopy data on crystals of  Ba$_{0.55}$K$_{0.45}$Fe$_2$As$_2$ show that in addition to a mid infrared band this material has a substantial metallic Drude band.  From the temperature and frequency dependence of this band we find that the charge carriers are coupled to a broad bosonic spectrum centered around 25 meV with a strongly temperature dependent coupling constant suggesting that magnetic excitations are responsible for the scattering. Comparing our data with the high temperature cuprates we see that iron arsenic compounds exhibit several similarities with the cuprates:  the bosonic spectral function is temperature dependent, evolving from a broad featureless band at high temperatures to peaked spectrum just as superconductivity sets in.  In the cuprates with $T_c=90$ K, in the underdoped region, a resonance appears around 40 meV in the middle of a gap of the spin fluctuation spectrum at ($\pi,\pi$) in reciprocal space.  Here we see a similar tendency of increasing spectral weight at 25 meV as the temperature is lowered.  It remains to be seen if these parallel tendencies survive future experimental tests with better crystals and a wider range of compounds.

\centerline{\bf Acknowledgments}

This work has been supported by the Natural Science and Engineering Research
Council of Canada and the Canadian Institute for Advanced Research.
Sample growth and basic characterization done at the Ames Laboratory was supported by the Department of Energy, Basic Energy Sciences under Contract No. DE-AC02-07CH11358.

\end{document}